# Density Gradients and Absorption Effects in Gas-filled Magnetic Axion Helioscopes

R.J. Creswick [1], S. Nussinov [1,2], and F.T. Avignone III [1]

[1] Department of Physics and Astronomy, University of South Carolina, Columbia, South Carolina, 29208, USA
[2] Department of Physics, Tel Aviv University, Tel-Aviv, Israel 69978

The effects of absorption in the gas, and of density variations on the sensitivity of gas-filled solar-axion helioscopes are theoretically investigated. It is concluded that the 10-meter long CAST helioscope, the most sensitive experiment to date is near the limit of sensitivity in axion mass. Increasing the length, gas density, or tilt angle all have negative influences, and will not improve the sensitivity.

## 1. INTRODUCTION

The axion, the Goldstone boson associated with the spontaneous symmetry breaking of the Peccei-Quinn Symmetry [1-3], could be generated in the core of the sun via Primakoff interactions with nuclear electromagnetic fields. The axions could similarly be reconverted to photons in magnetic fields perpendicular to their velocities and detected with photon detectors at the end of a magnetic helioscope. Both processes are driven by the Primakoff diagram via the Hamiltonian,

$$\mathcal{L}_{a\gamma\gamma} = a\vec{B} \cdot \vec{E} / M \qquad (1)$$

In (1), $\vec{B}$ and $\vec{E}$ are magnetic and electric fields, respectively, $a$ is the axion field, and $M \equiv 2\pi F_a / \alpha$, where $F_a$ is the scale of the spontaneous symmetry breaking. For a value of $F_a \approx 10^7 GeV$ ($M \approx 10^{10} GeV$), and $m_a << 1 keV$, for example, one expects a solar axion flux of the order of $\Phi_a = 2.1 \times 10^{11} / cm^2 \cdot s \cdot keV$ with a peak in the spectrum at about $4 keV$ [4].

In 1983, P. Sikivie introduced the concept of the magnetic helioscope [5], and since then there have been three well known solar-axion searches with this technique: the Brookhaven experiment in 1992 [6], the Tokyo experiment in 1998 [7], and the most sensitive one, the CAST experiment currently operating at CERN [8]. CAST has set a lower bound $M \geq 10^{10} GeV$ for very low-mass axions. However, to be sensitive to axions with $m_a \geq 10^{-2} eV$, a cold gas is introduced into the bore of the magnet to slow the photons to maintain coherence of the axion and photon wave functions. As the magnet is tilted to track the sun, the gas density varies along the length, and changes as the magnet continues to rotate. The purpose of this work is to determine the effects of these conditions on the sensitivities of magnetic helioscope solar axion searches.

For axion masses on the order of an electron volt and $E_a \sim 4 keV$, $qL >> 1$ where $q = m_a^2 / 2E_\gamma$ is the wave vector for the $a \rightarrow \gamma$ oscillation and $L$ is the path length in the magnet. For example in CAST, $L = 9.3m$ and $qL \sim 10^4$. Coherent regeneration of the photon is then limited to distances of the order of a centimeter, greatly reducing the expected signal.

To restore coherence between the photon and the axion, the bore of the magnet is filled with a buffer gas. The plasma frequency acts as an effective mass of the photon

$$\omega_p^2 = \frac{4\pi n_e e^2}{m_e} \equiv m_\gamma^2 \qquad (2)$$

where $n_e$ is the total electron density. In the presence of a buffer gas, $q$ becomes

$$q = \frac{m_a^2 - m_\gamma^2}{2E_\gamma} \qquad (3)$$

When $q = 0$ the axion and photon remain in phase over the entire length of the magnet, leading to a conversion rate that scales as $L^2$. One can estimate the width of the resonance (as a function of the axion mass) by setting $qL = 2\pi$, which gives

$$\Delta m_a \sim \frac{2\pi E_\gamma}{L m_\gamma} \qquad (4)$$

For values of the parameters typical of CAST and $m_\gamma = 1.0 eV$, $\Delta m_a \sim 1.6 \times 10^{-4} eV$, which illustrates how narrow the ideal resonance is. The response of the CAST helioscope is shown in figure 1.

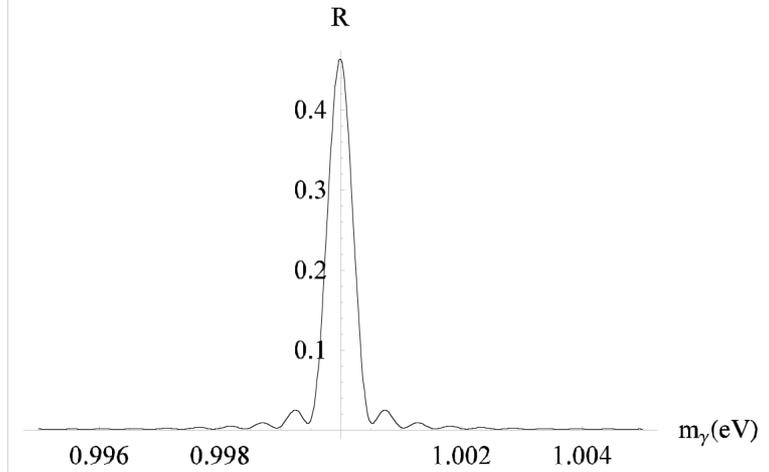

Figure 1. Relative axion to photon conversion probability in the CAST helioscope for $m_a = 1.0 eV$, $E = 4.0 keV$, and $\theta = 0°$.

However, absorption of the photon through the photoelectric effect and variation in the plasma frequency due to density gradients when the magnet is tilted away from horizontal will affect the counting rate. Specifically, the peak of the resonance will be reduced and there will be an asymmetry in the counting rate as a function of the electron density. In the following we discuss this in some detail.

## 2. AXION TO PHOTON CONVERSION BY THE PRIMAKOFF PROCESS

The evolution of the photon amplitude, $A(z)$, along the bore of the magnet $0 \le z \le L$ is given by

$$\frac{\partial A(z)}{\partial z} = \left(iq(z) - \frac{\Gamma}{2}\right)A(z) - i\frac{B}{2M}a(z) \qquad (5)$$

where $a(z)$ is the axion amplitude, $\Gamma = n_G \sigma_G$ is the absorption coefficient, which we take to be constant, and

$$q(z) = \frac{m_a^2 - m_\gamma^2(z)}{2E_\gamma} \qquad (6)$$

Since $B/2M \ll 1$ the axion amplitude can be taken to be constant over the whole path, $a(0) = a(L) = 1$, and (3) can be integrated at once, giving

$$A(z) = -\frac{iB}{2M} e^{i\int_0^z dz'(q(z')+i\Gamma/2)} \int_0^z dz' e^{-i\int_0^{z'} dz''(q(z'')+i\Gamma/2)} \qquad (7)$$

$He^3$ is chosen as the buffer gas in both CAST and the Tokyo experiment because it does not liquefy at 1.9K, the operating temperature of the magnet, and (apart from hydrogen) has the smallest photoelectric cross section. At energies $E_\gamma \sim 4keV$, $\sigma_{He^3} \sim 4.\times 10^{-24} cm^2$ and increases very rapidly as the energy is lowered. The nominal $m_a \sim 1.0eV$ is matched by an electron density $n_e = 2n_{He} = 7.28 \times 10^{20} cm^{-3}$ which gives $\Gamma(4kev) \sim 1.6 \times 10^{-3} cm^{-1}$.

When the CAST magnet is at the maximum tilt angle of $\theta_0 = 8°$, the end-to-end height difference is 1.3m. The gas density varies along the bore according to an isothermal Boltzmann distribution,

$$n_{He} = n_0 e^{-mgz\sin\theta/kT} \qquad (8)$$

and the end-to-end fractional density variation is

$$\delta n/n = m_{He} gL\sin\theta/kT \qquad (9)$$

which, under the conditions of the CAST experiment, gives $\delta n/n \sim 2.45 \times 10^{-3}$. If the gas density is "tuned" so that $q(z = L/2) = 0$, then

$$q(z) = \frac{m_a^2}{2E_\gamma} \frac{m_{He} g(z - L/2) \sin\theta}{kT} \qquad (10)$$

The total end-to-end variation in $q$ for values typical of CAST and $m_a \sim 1.0 eV$ gives

$$\Delta q = \frac{m_a^2}{2E_\gamma} \frac{\delta n}{n} \sim 1.56 m^{-1} \qquad (11)$$

which implies that the photon and axion get out of phase in a distance on the order of 2.0m.

In general the relative wavevector between the photon and the axion is, in terms of the rescaled coordinate $z \to z/L$

$$q(z) = q_0 + (z - \tfrac{1}{2}) b \qquad (12)$$

where,

$$b = 14.4 \left(\frac{L}{10m}\right) \frac{m_\gamma^2}{eV^2} \frac{4keV}{E_\gamma} \frac{\sin\theta}{\sin\theta_0} \qquad (13)$$

and

$$q_0 = 6.0 \times 10^3 \frac{(m_a^2 - m_\gamma^2) eV^2}{E_\gamma / 4keV} \qquad (14)$$

allows for the possibility that the photon mass may not match the axion mass at the center of the helioscope.

In these same units the absorption coefficient, $\Gamma$, is given by:

$$\frac{\Gamma}{2} = 0.8 \left(\frac{m_\gamma^2}{eV^2}\right) \frac{\sigma_{P.E.}(E)}{\sigma_{P.E.}(4keV)} \frac{L}{10m} \qquad (15)$$

We can define the suppression of the signal as the ratio of the conversion probability to that at resonance and in the absence of absorption and variation in the gas density,

$$R = e^{-\Gamma} \frac{\pi}{2b} \left| e^{ia^2/2b} \left\{ \mathrm{erf}\left[\sqrt{\frac{i}{2b}}(a+b)\right] - \mathrm{erf}\left[\sqrt{\frac{i}{2b}} a\right] \right\} \right|^2 \qquad (16)$$

where, $a = (q_0 - b/2) + i\Gamma/2$.

Figure 2. Relative conversion probability as a function of $m_\gamma$ for a tilt angle of 8° and $m_a = 1.0 eV$. The slight shift in the peak from $m_\gamma = 1.0 eV$ is due to the effect of absorption.

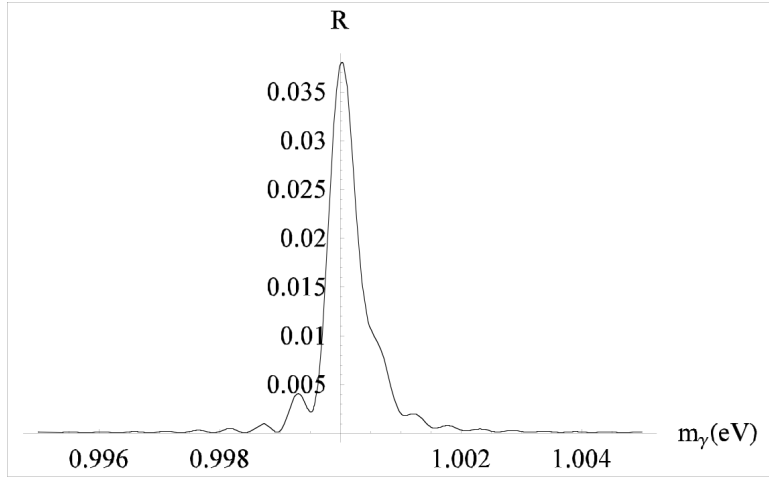

Figure 2 shows the ratio, $R$, as a function of the photon mass, $m_\gamma$, for a tilt angle of 8°, $m_a = 1.0 eV$, and nominal values for the other parameters.

Variable density along the path is *not* equivalent to sampling different axion masses. To correctly achieve the latter, we need to repeat the experiment with several different uniform densities so that one value of $m_a$ is searched for in each run. In the present arrangement the axion-photon system encounters a range of $q$-values and densities along the path. Together with absorption this reduces the expected photon signal and shifts its spectral distribution to higher energies.

The interplay of these two effects also produces an amusing asymmetry between the case where the Sun is above the horizon and when it is below. Suppose the density of the buffer gas at $L/2$ is set a little above the resonant value for the proper axion mass. Then resonance will occur at some point in the upper half of the telescope. If the Sun is above the horizon, axions enter the upper end of the telescope, transform into photons and, on average, travel a distance greater than $L/2$ to the detector at the opposite end. On the other hand, if the Sun is below the horizon, the axions enter the lower end, and travel a distance less than $L/2$ to the detector at the other end. Because of absorption, the signal in the first case will be reduced relative to the signal in the second case. If the central density is slightly below the resonant value, the sign of this asymmetry is reversed.

## III SUMMARY AND CONCLUSIONS

The variation of the density of the buffer gas along the length of the telescope and absorption of the photon both reduce the expected signal and hence the sensitivity of the experiment. If, however, we correctly account for the dramatic decrease in $\Gamma$ and

$q_0 \sim E_\gamma^{-1}$ with energy, this may be somewhat compensated. Also, the correlation of the asymmetry of the signal with the position of the Sun can be used to distinguish positive and negative detuning of the central density.

On the whole, however, the above analysis suggests that insofar as searches for solar axions using buffer gases, CAST in its present design faces some ultimate limitations. Absorption effects mitigate against increasing the length of the magnet. Increasing the maximum tilt angle would allow tracking the Sun for a longer time, but this would result in increased effects of detuning along the length of the magnet. Both absorption and detuning increase as $m_a^2$, which severely limits the sensitivity of to axions with masses more than a few $eV$. However, searches for axions much lighter than 1 eV, where vacuum coherence is maintained without buffer gas, are nor affected by the above analysis.

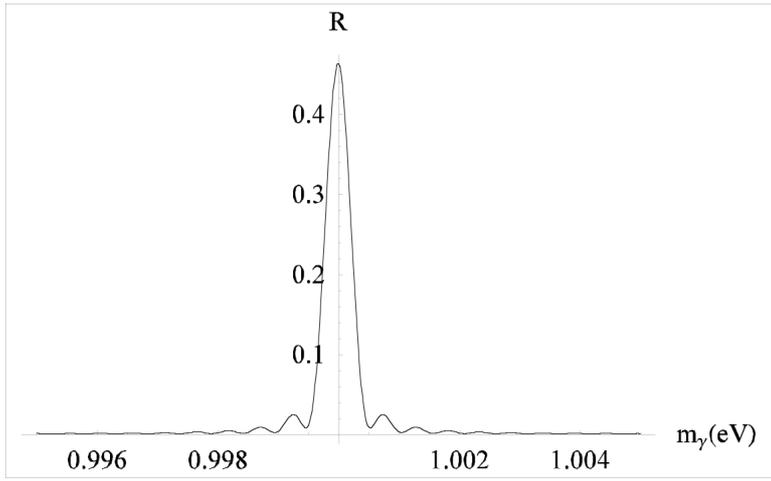

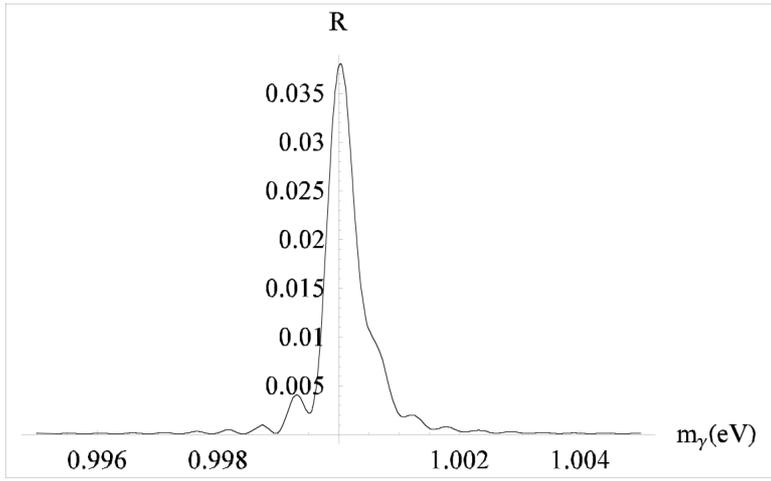